# EPISTEMIC CONTROL AND THE NORMATIVITY OF MACHINE LEARNING-BASED SCIENCE

Emanuele Ratti[1]

Department of Philosophy, University of Bristol

**Abstract**. The past few years have witnessed an increasing use of machine learning (ML) systems in science. Paul Humphreys has argued that, because of specific characteristics of ML systems, human scientists are pushed out-of-the-loop of science. In this chapter, I investigate to what extent this is true. First, I express these concerns in terms of what I call 'epistemic control'. I identify two conditions for epistemic control, called 'tracking' and 'tracing', drawing on works in philosophy of technology. With this new understanding of the problem, I then argue against Humphreys' pessimistic view. Finally, I construct a more nuanced view of epistemic control in ML-based science.

**Keywords**: machine learning; epistemic control; cognitive values; normativity

1. INTRODUCTION

In the past few years, there has been a growing interest in the challenges of keeping meaningful control over automated computational tools. The issue of so-called 'human-in-the-loop' has been raised especially in the contexts of autonomous weapon systems (Human Rights Watch 2012), public institutions (Danaher 2016), and credit scoring systems (Citron and Pasquale 2014). The issue is raised because such automated processes are so complex, large-scale, and fast, that humans cannot simply monitor them in ways in which other tools are monitored. More precisely, this problem can be defined as the extent to which humans can keep control over automated tools by deciding when to use them, or preventing that these tools are used in a certain situation, or overseeing how the tools take decisions in specific instances (and, potentially, in all instances). For example, Danaher says that in the case of data mining systems humans "can predetermine the patterns that data-mining algorithms search for (…) can review and scrutinize the recommendations made by algorithms, or they can essentially leave it up to the machines" (2016, p 248).

An unexplored side of this debate is that such automated computational tools (and in particular AI tools) are used in science as well, and hence the same issues of 'human-in-the-loop' and human control potentially apply in the context of scientific practice. Where scientific processes are automated, we must make sure that there is proper human oversight to ensure that scientific practice is not led astray by poorly designed or poorly adapted tools. While in the cases mentioned above the payoffs of having humans in the loop are especially ethical and legal, in the case of scientific practice the payoffs are also epistemic. I call this effort of keeping humans-in-the-loop of science 'epistemic control', which I will define more precisely in Section 2.

---

[1] mnl.ratti@gmail.com

The aim of this chapter is to analyze the relation between the use of machine learning (ML) in science, and human epistemic control. Asking the question of epistemic control in ML-based science is important, given the increasing importance of ML tools in scientific practice, and the control-related issues that have been emerging in this context. In particular, recent analyses of the opacity of ML systems have raised concerns over the unpredictable ways in which ML shapes scientific methodologies and scientific outputs (Creel 2020; Boge 2022). Latest implementations of large language models in fields like biology or medicine (Mesko and Topol 2023; Bommasani et al 2021) have shown the challenges of controlling the content of the data sets used, how they are processed, and how the models are transferred across different contexts.

The structure of the chapter is as follows. In Section 2 I will introduce the notion of 'epistemic control' and I will reconstruct the argument for the lack of epistemic control in ML-based science. This is based on Paul Humphreys' work that, I think, represents the most comprehensive and systematic attempt at making the claim that ML-based science can potentially be 'out of human control'. Humphreys' argument is based on a fine-grained analysis of how ML algorithms process data and the opacity of the learned representations. In Section 3, I will first highlight two weaknesses in Humphreys' argument (3.1 and 3.2), showing that there is more to ML-based science other than the aspects he emphasizes, and that scientists can keep meaningful epistemic control on other important components of ML systems that Humphreys neglects. Finally, I will complement the analysis laid out in 3.1 and 3.2 by describing the two-way dynamic underlying epistemic control in ML-based science, specifically how the use of ML in science can lead scientists to commit to certain methodological choices that are subtle, but nonetheless have profound epistemic consequences, especially for the aims of science (3.3). I call this 'the normativity' of ML-based science.

## 2. EPISTEMIC CONTROL IN MACHINE LEARNING-BASED SCIENCE AND HUMPHREYS' PESSIMISTIC VIEW

## 2.1 Epistemic Control

I define the concept of epistemic control by adapting the notion of 'meaningful human control' (de Sio and van den Hoven 2018) to the scientific context. Debates on this notion start with a simple question concerning autonomous weapon systems: who can be said to be responsible for the wrongdoings of autonomous weapons systems? For humans to be responsible, it is said, they have to be ultimately in control. But for 'being in control', mere 'human-in-the-loop' is not a sufficient condition, because just being in the loop does not imply being able to influence the system or understanding it, and these are additional conditions that intuitively are important for control. For this reason, some have proposed the idea of 'meaningful human control'. According to de Sio and van den Hoven (2018), meaningful human control requires two conditions. The first is called 'tracking', namely that a system is under meaningful human control when the system's actions track relevant human moral reasons, in the sense that the system is demonstrably and verifiably responsive to the relevant human moral reasons. The second condition is called 'tracing': to be under meaningful human control, there should be tracing relations between the mechanism of decision-making of the system, and the technical and moral understanding of some relevant humans involved. It is important to point out that their account is tailored for systems where many human agents are involved in their construction and implementation, and as long as for some of these agents the two conditions apply, then there is meaningful human control. My conception of epistemic control draws significantly from de Sio and van den Hoven's account, but it differs in two respects.

*First*, the two contexts are different. For instance, de Sio and van den Hoven are especially interested in the moral dimension of control, while my interest is epistemic. Moreover, the boundaries of the system to be under meaningful control are sharper in de Sio and van den Hoven's case, because it is about sharply identifiable systems (i.e., a drone or a tank). In the case of science, it can be something as broad as a scientific project, something more specific and concrete as a scientific instrument, an experimental system, or something abstract like a methodology. I subsume all these possibilities under the label 'scientific item[2]'. A *second* difference is that it is not clear whether de Sio and van den Hoven's account is discrete or not. But in my account, there is a continuum of control that human scientists can exercise over the scientific endeavor. On these bases, I propose to modify the two conditions in the following way to adapt them to an epistemic context like science:

---

[2] This is vague and unsatisfactory, but for reasons of space, I will develop this idea in another work

- *Tracking condition*: a scientific item must be responsive to scientists' epistemic or methodological standards/criteria that are relevant in a given context; what a scientific item does should track or be aligned to epistemic or methodological criteria or standards that a community of scientists accept. These 'criteria' or 'standard' are discipline-specific. For instance, in the case of a standard polymerase chain reaction (PCR), 'tracking' is obtained by making sure that each step of the process (i.e., denaturation, annealing, elongation, detection) is responsive and aligned to latest, updated standards and benchmarks created by the biological community. Tracking conditions are about establishing whether a scientific item indeed aligns to given scientific standards. If one cannot establish that, then tracking conditions do not hold
- *Tracing condition*: one or more scientists must be able to trace, from a given methodological and epistemic perspective, how the scientific item works or unfolds; these 'perspectives' are, again, discipline-specific. Going back to the example of the PCR, tracing conditions apply when one or more scientists are able to tell how the final outcome of the PCR has been obtained by appealing to methodological and epistemic criteria accepted by the biological community. Tracing conditions are about understanding the behavior of a scientific item through discipline-specific standards. If one cannot understand how the behavior unfolds, then tracing conditions do not hold

There is meaningful human epistemic control (for short, *epistemic control*) when these two conditions apply. But, as I said above, there is a continuum of control. Therefore, a more precise statement would be to say that *epistemic control grows when the degree of instantiation of the two conditions increases*.

Defined in this way, epistemic control is an ideal that has been pursued in various ways in the history of science under different names. Prominent exemplifications include attempts at mechanizing scientific methods or providing grounds for 'procedural objectivity' (Nickles 1980; Gillies 1996; Megill 1994); replacing idiosyncratic personal judgements with impersonal standards (Davis 2019; Porter 1995); organizing a collective process of oversight over scientific work (Longino 1990). In these cases, tracking and tracing conditions have been implemented in various ways: by establishing how to evaluate scientific items, it is possible to tell whether these align with standards established by the community; by mechanizing scientific methodologies, scientific procedure are transparent in such a way that one can ascertain whether they are responsive to accepted epistemic and methodological standards, and the community of scientists will be able to trace them; by replacing idiosyncratic personal judgements, we make sure that subjective aspects, which have been typically seen as not responsive to methodological and epistemological criteria accepted in the community and are implicit and tacit (i.e. they cannot be tracked and traced), are kept to a minimum. It is also important to add that epistemic control does not imply epistemic success, because a scientific item might be easy to track and trace, but then one might find that it is simply not aligned with the accepted methodological and epistemic criteria. However, having more epistemic control facilitates the process of scrutiny which will increase the chances of at least spotting problematic scientific items.

In which sense in ML-based science there can be a lack of epistemic control? I will expand on this below, but for the time being let's look at two examples that illustrate the intuition behind losing epistemic control in ML-based science. These examples refer to errors that are peculiar to ML. Consider shortcuts first. These are errors that happen when a ML system achieves high predictive accuracy by leveraging features of data sets that have little to do with the target phenomenon the ML system is supposed to learn from. For instance, a deep neural network (DNN) might classify medical images as 'cancer/non-cancer' on the basis of non-medical information erroneously encoded in the image itself. A second example are adversarial attacks, which is when target manipulations of data sets (that can be imperceptible to humans, such as modifying a few pixels in images) lead to a drop in performance. These are ML-specific errors that we want to avoid. But in both cases, it is difficult for human scientists to anticipate these issues. It is difficult to anticipate shortcuts, because often what the ML system leverages to make predictions cannot be neatly identified or interpreted through meaningful human concepts (Boge 2022). In such cases, one cannot establish that the outputs are responsive to or aligned with relevant discipline-specific epistemic and methodological standards (e.g., classifying an image as a 'cat' because of cat-related pixels): the tracking condition does not hold. Anticipating adversarial attack is challenging because of the complexity of the algorithmic processes (Freiesleben and Grote 2023) – it is difficult to establish *tracing relations* between the way ML system unfolds and the relevant methodological understanding.

These problems have suggested that scientists might have little control over how ML systems learn or their outputs. But beyond considerations about the complexity of computational systems, a precise account of how the characteristics of ML concretely undermines epistemic control is hard to find. The only exception is Paul Humphreys' work (2011; 2020; 2021). His concerns about ML in sciences get close to the idea of losing epistemic control – and this is why I will focus on his writings to get a baseline understanding of the relation between ML and epistemic control. In order to understand his ideas and how they are connected to my concept of epistemic control, we need to understand two components of his view; first, the view that computational systems brings a new 'epistemic perspective' in science; second, that this epistemic perspective is irreducible to a human perspective because of opacity.

**2.2. Epistemic Perspectives**

Let me start with 'epistemic perspective'. According to Humphreys (2011), one interesting aspect of computational science is that "it uses methods that push humans away from the centre of the epistemological enterprise" (p 133). This, he continues, is not necessarily novel; as a matter of fact, many scientific instruments have 'pushed away' humans, at least when perceptual tasks were concerned. But in the case of computational sciences, the push is much more radical: it is a divorce. Science has always been anthropocentric, Humphreys says, but now the situation can change. This divorce is a radical displacement that can happen, according to Humphreys, because computational systems are akin to a new scientific authority. This might potentially lead to "abstract[ing] from human cognitive capabilities when dealing with representational and computational issues" (Humphreys 2011, p 134). But what does Humphreys exactly mean by 'new scientific authority'?

This is the way I understand Humphreys' views. He says that the idea of an exclusively anthropocentric epistemology may not be suitable to describe what is currently happening in science. In (2021), he sketches a notion of 'machine knowledge', as something that can characterize a machine's point of view on what can be known. In (2020), he also refers to patterns that machines can discover to which humans are not accustomed, or simply unable to interpret. Humphreys' idea is that there is something like a point of view on the objects of scientific knowledge (e.g., natural phenomena) that is unique of computational tools such as ML or computer simulations; and this point of view may be problematic for human beings, in the sense of something that humans cannot understand it. This 'problematic' point of view on the object of knowledge constitutes, according to Humphreys, a 'new scientific authority'.

I call these views associated to point of view/scientific authority 'the epistemic perspective of ML'. To be fair (and to my knowledge), Humphreys does not use the term 'perspective'. He speaks at times of 'style of reasoning', and he refers explicitly to Crombie's and Hacking's works. But he ended up rejecting the notion, because 'reasoning' has an anthropocentric flavor (2011, p 133). He also uses the expression 'styles of representation' (2020). These are unique ways of representing phenomena (from a specific angle and not another), and they include specific ways of manipulating these representations to construct knowledge (i.e. techniques). But Humphreys' view sometimes seems to encompass much more than just representations, including epistemic interests, problems and questions, hypotheses, theories and models, presuppositions, methods. For this reason, 'epistemic perspective' might be a better term, exactly because this typically encompasses all those aspects mentioned just above (Plumacher 2012). The notion of epistemic perspective plays both an analytical function (i.e. it determines the epistemic object and its conditions), and an ordering function (i.e. it helps us navigating epistemic practices), which cannot be adequately covered by 'styles of representation'.

The issue with respect to the human's and the computational tools' perspectives boils down to one specific question: "are there distinctive styles of representation [i.e. epistemic perspectives] that are involved in automated science [i.e. computational science]" (2020, p 13) which involve nonhuman agents? Especially when computational tools like DNN are concerned, Humphreys argues that this is indeed the case. Even though hardware, software, and architectures are built by humans, still "the details of the internal representations are constructed by the computer, not by humans" (p 22). These internal representations constructed by the machine constitute a significant part of the perspective of computational tools. This 'machine perspective', according to Humphreys, should be neatly distinguished from the perspective of human scientists.

Humphreys does not say what the human point of view or 'human perspective' really is. In the case of science, we can say that the 'human perspective' is discipline-specific: it is constituted by the given discipline-specific norms, the background knowledge including an ontology underlying the given scientific field (which can be used to 'build' discipline-specific scientific representations), and specific research questions and goals[3].

Humphreys says that this new computational perspective can displace or replace the human perspective. Why? In what he calls the 'hybrid scenario' (a scientific scenario where humans and machines coexist), humans' and computational tools' perspectives are kept close, by 'translating' the relevant parts of computational procedures and outputs into the 'grammar' of a specific scientific human perspective. This means that there is push to subsume the tool's perspective onto the human's perspective. As a consequence, human scientists can track outputs and trace what a given computational tool is doing on their own terms, and evaluate and/or amend the conduct of the tool itself on the basis of methodological and epistemic criteria

---

[3] See my paper (2020) for how this would work in a specific discipline (i.e., molecular biology)

that are accepted by a given community of human scientists. However, according to Humphreys some computational systems do not allow this translation – and ML is the most prominent case.

**2.3 Opacity**

Humphreys makes the convincing point that the 'translation' is sometimes not possible because of the phenomenon of *(essential) epistemic opacity*.

In the context of computer simulations, Humphreys says that a process is (essentially) epistemically opaque "relative to a cognitive agent X at time t (…) if and only if it is impossible, given the nature of X, for X to know all of the epistemically relevant elements of the process" (2011, p 139). This preliminary characterization refers especially to computer simulations. How does Humphreys adapt this conception to ML[4]?

Humphreys characterizes the opacity of ML systems like DNN as 'representational opacity' (2021). The 'representation' is the actual content of the model that the DNN system builds. But this 'representational opacity', in the way Humphreys talks about it, refers to more than just the content of the model. In particular, it is about two things.

First, we do not know much about how DNNs construct their models. While at a basic level DNN is a function approximation device, functions between layers are nonlinear and complex to the extent that no human can really interpret nor understand what is going on. We do know the forms of the generic architectures that DNNs can possibly take, but more relevant details are beyond our reach. For instance, Humphreys mentions that we do not know the relevant details of the algorithm for updating during the training procedure. But the opacity can also lie in modeling activities in the ML pipeline that are taken over by the ML system itself, and that were typically done by humans. Take feature engineering, which is a central procedure for well-functioning ML systems. This is a manual process done by human scientists to extract derived features from raw features of data sets, and that will be used as inputs for training procedures. However, there is an interesting trend towards automating feature engineering by turning it into an optimization procedure like the one leveraged for weight-parameters learning (Termine et al 2026). It has been shown that one can extract useful derived features by using convolution, which is a linear transformation based on the application of a kernel of parameters to the input. What happens is that the various regions of the input are processed through a filter computing the weighted sum of the input-features (i.e., the pixels of the image) in the region and kernel parameters, thereby mapping the result into a feature map. The high-level features produced are called 'embeddings', which the ML system uses, as it happens for human-constructed features, for predicting the target. In a field like psychiatry (Eitel et al 2023), typically, a psychiatrist will help ML practitioners to craft hand-made high-level features on the basis of what is relevant to psychiatry as a field. For instance, one can select from images high-level variables like cortical thickness, fractional anisotropy, etc, then determine their values from raw data, and finally end up with new features. This will facilitate tracing, because this process can be monitored and explained by appealing to discipline-specific standards and criteria. But in the case of machine-constructed features, tracing is challenging because it is difficult to understand how the features are constructed, and whether these procedures reflect discipline-specific considerations, unlike in the case of hand-made features.

---

[4] The literature on opacity has exploded in the past few years. Landmark contributions include (Creel 2020; Boge 2022; Sullivan 2022). Given that here I am engaging with Humphreys' view, I will only discuss his ideas on opacity. This does not mean that other views on opacity are not important.

The opacity lies also at the level of the representation itself, as Humphreys stresses. He characterizes DNN representations as implicit and distributed. When a representation is *implicit*, we need inferences or transformations to identify the referential content of the representation, but often in DNN "the methods for making the required inferences are unknown" (p 12). This is reflected also in machine-constructed features: embeddings might represent magnitudes of the input that do not possess any clear meaning for the scientists in the relevant domain. In (Alvarado and Humphreys 2017), it is argued that opaque representations "have features that do not correspond to familiar linguistic concepts, [and] we are faced with the questions of whether some of those representations are permanently unknowable" (pp 742-743). DNN's internal representations may refer to aspects of reality, though it is entirely possible that these are aspects "to which humans are not accustomed, and perhaps not even able to interpret" (2020, p 22). In turn, this implies a lack of epistemic control over outputs, because no one is able to tell how the model 'represents' or 'accounts for' a phenomenon of interest. The fact that human scientists might not be accustomed to ML's representations, depends also on the fact that these are *distributed*. This has to be contrasted with 'local', which is when most elements of a model (e.g., variables, parameters, etc) represent an element of the phenomenon that the model is targeting (where this is mediated by background scientific knowledge). Scientific models are, often, local (or element-wise representational as Freiesleben et al 2024 say), and components are characterized through the lens of well-defined terms and concepts which refer to relevant aspects of natural phenomena[5]. But in the case of DNNs, a feature can be represented by many nodes, and each node participates in representing many features. This means that the representation is 'distributed' (Freiesleben 2025), unlike in traditional scientific models where representations tend to be 'local'. How a human scientist interprets this 'distributed representation' in light of the typical scientific categories used within a context remains an open question – this means that the 'tracking' of outputs (in this case, ML models) often fails.

**2.4 What about epistemic control?**

To summarize, let me say more explicitly how considerations about epistemic perspectives and opacity, in the way understood by Humphreys, are connected to my notion of epistemic control. The reasoning flows as follows. First, ML provides a new *epistemic perspective*: there is a new point of view on the object of knowledge belonging to ML systems, which emerge in particular in the characteristics of the representations constructed, as well as in the way these are constructed. This new perspective is *opaque* in the way indicated above, and this opacity challenges the process of 'translation' of this new perspective into human's perspective. This *leads* to a lack of epistemic control: the opacity of how DNNs construct their models and what their models 'represent' decreases human's ability, within a given scientific context, to *track* to what 'reasons' these systems are responsive to, and it makes difficult to *trace* their rationale.

**3. A BALANCED VIEW ON EPISTEMIC CONTROL IN ML-BASED SCIENCE**

Does ML-based science really undermine epistemic control in the way Humphreys has argued? In this section, I argue that epistemic control is not impaired in exactly that way. I start first by highlighting two problems with Humphreys' view (3.1 and 3.2). Next, I propose a more balanced view of how epistemic control relates to ML-based science. I will say that scientists can avoid the misplacement Humphreys (i.e. the total loss of epistemic control) talks about (3.1 and 3.2). This is because the 'perspective' of ML-based science is wider than just ML internal representations; in fact, many components of ML systems are shaped in light of human scientists' own cognitive values, which limit the issues raised by essential epistemic opacity. However, the human perspective – and epistemic control - is also constrained by what I call the 'normativity' of ML-based science (3.3), which is the fact that the use of ML in science can shape the range of desiderata of the discipline to which ML is applied to.

---

[5] A caveat is important. There might be components that do not refer to anything specific or refer to something unknown. This is the case of, e.g., 'noise'. Models are also characterized by idealizations, and some components might not refer to things that 'exist' (e.g. infinite populations).

**3.1 ML-based Science and Opacity**

The first problem with Humphreys' view is about the extent to which opacity stands in the way of the epistemic control required to achieve specific scientific goals. While it might be true that opacity diminishes epistemic control in the way highlighted above, the type of control it impairs is problematic only for certain goals, and not others. Let me unpack this, specifically for the problem of the opacity of representations[6].

In (Lopez-Rubio and Ratti 2021), I have shown that in cancer genomics ML models can be opaque, in the sense that they cannot be interpreted according to the explanatory criteria used for biological models (i.e. the criteria of mechanistic explanations). For a mechanistic model to be explanatory, it has to layout the organization (i.e. 'causal connectivity') between relevant entities and activities involved in that particular mechanistic phenomenon. Pinpointing the organization is done in various ways, but in general it is a procedure that requires the model to be intelligible, which is understood in the mechanistic literature in terms of 'usability' or 'built-it' tests (Craver and Darden 2013). A corollary of these considerations is that the more a mechanistic model is complex in terms of number of entities involved, the more it will be difficult for us to grasp the organization between them, and hence the model will be less and less explanatory. ML models perform better when the number of features increases[7] – and this increases the size of models too: good ML models are large, and this clash with our norms concerning the construction of mechanistic explanations. This is because in order to construct a mechanistic explanation out of a ML model, we need (a) to identify the components of the ML model and to what components of possible explanations (e.g. biological entities and processes) these might be associated to, and (b) understand the relations between the components of ML model in order to 'translate' these into appropriate relations between the components of possible explanations (i.e. the organizational aspects of mechanistic models). This is, in theory, how we 'track' ML models and align them with explanatory standards. But, as I show in my paper with Lopez-Rubio (2021), this is an impossible task. Therefore, using

---

[6] The argument addressing the opacity of the way representations are constructed is analogous, but I do not have enough space to cover it here. Therefore, my discussion will relate only to tracking conditions

[7] This is a classic consideration that was made by pioneers of ML. For instance, in commenting SVM, Breiman notices that in order to increase the chances that a separating hyperplane exists, the trick is to add more features – "[t]he higher the dimensionality of the set of features, the more likely it is that separation occurs [up to a point where overfitting and computational complexity become problematic, though]" (2001, p 209)

better performing ML models in biology decreases our ability to evaluate and learn from their representations, where this impairs especially tracking conditions. What happens is that we might not be able to 'track' how the relations between features that the ML models has identified, correspond to relations between biological entities and processes that can be represented as a neat and well-organized mechanism: in other words, biologists cannot establish how ML model's results relate to the explanatory standards used in biology, leaving them unable to evaluate whether the ML model is responsive to domain-related criteria. But while this lack of epistemic control is real, *it is still relative to a certain goal, and in particular explanatory goals*. If we change goal, considerations of epistemic control change too. In particular, what we need to 'track' will depend especially on the goal that we have in mind. In cancer genomics, explaining cancer mechanistically is one goal among many. Another important goal is, for instance, molecular stratification of cancer. In this case, the task of classifying does not depend on whether or how the relations between components of ML models can be translated into an organizational picture sensitive to the criteria of mechanistic explanations. The ML system will just classify tumors in different classes. In order to say whether the ML model is responsive to discipline-specific standards (i.e., tracking conditions), classification outputs will be evaluated by comparing them to a number of available clinical and laboratory measures, to make sure that they are coherent and really grasp stable conditions. Therefore, for the sake of classifying with ML models, tracking conditions are available. This means that epistemic control is not impaired.

Similar points have been made recently, especially concerning the relation between opacity, accuracy, and explanations (London 2019). However, they are not new at all. In fact, they go way back to early debates on the goals of ML, understood as a distinct type of statistical modeling. For instance, Breiman (2001), the inventor of random forests, characterizes the 'algorithmic modeling culture' (which significantly overlaps with ML) in opposition to a 'traditional' way of doing statistics. While more 'traditional' statistics leads often to "a simple and understandable picture of the relationship between the input variables and the response" (p 203) and this is its goal, in algorithmic modeling the goal is simply to find a function that predicts well, independently of what the underlying model is telling us about the relations between variables and - I would add, without distorting Breiman's views- of whether the information contained in the model can be organized into an explanation. Vapnik (2006), one of the creators of SVM, explicitly connects ML to an instrumentalist perspective, emphasizing that the important goal is prediction (and generalization). In his view, understanding what these models tell us about the world and, as in the case of Breiman, their explanatory nature, are not central aspects. Assuming we have ways to establish whether the predictions are aligned to certain standards established by the relevant scientific community, epistemic control is not impaired.

To sum up, Humphreys' idea that because of opacity humans are pushed away from the epistemological center of ML-based science may have been overstated, as opacity only impairs epistemic control related to a narrow set of scientific goals.

**3.2 ML-based Automated Science is Enveloped in Human Values**

A second problem with Humphreys' conception is that opacity characterizes only a part of a given ML system. As highlighted above, sometimes Humphreys identifies the 'machine' perspective with something quite broad. However, in most cases, Humphreys sees opacity as an unsurmountable problem because the entire perspective of ML-based science is reduced to ML internal representations and how these are built by the ML system itself. But if we explicitly take the notion of epistemic perspective on board, then representations are just one dimension of the 'point of view' of ML-based science. If we think about epistemic perspectives as encompassing epistemic interests, problems and questions, hypotheses, theories and models, presuppositions, methods, etc, then we should consider not just how ML systems build their representations, but also how systems themselves are built *by humans*.

Consider what it takes to build a ML system. A typical pipeline for training and deploying a ML tool in science (Figure 1) would go through an initial phase where the scientific problem to solve is defined and adapted to the ML context; a phase where data are acquired, and curated; another phase where models are developed, validated, and interpreted, and finally a phase of deployment. The shape of a ML system will not depend solely on the internal representations (most likely making an appearance only in the phase of model development); rather, design and training choices constitute significant factors in shaping how ML systems look like. For the most part it is the human perspective that molds ML systems, as it is the human subject who shapes systems by making specific design and training choices.

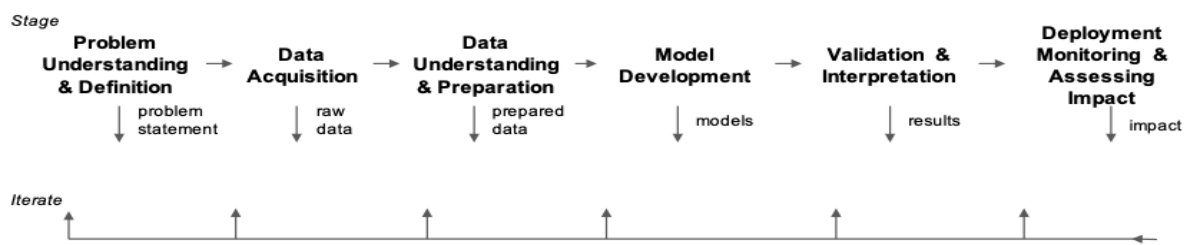

Figure 1. An idealized pipeline to build ML systems. Adapted from (Ratti and Graves 2021)

Saying that algorithmic systems are 'made' by human subjects implies that the technical choices made to build these systems are *value-laden*. Technical choices are value-laden because they tend to promote certain characteristics of the systems we are building that we think are desirable or valuable. With this I mean not only that moral and political values shape the training of algorithm. That these non-cognitive values are relevant in algorithmic systems is a widely shared view in all disciplines investigating data science tools, and their problems have been thoroughly investigated also in philosophy of science (Biddle 2020; Johnson 2021). What I am interested here is that, in addition to non-cognitive values, the procedure to construct algorithmic systems is characterized by technical choices that are motivated by *cognitive values*. The analysis of how cognitive values impact the construction of algorithmic systems has been largely neglected[8]. This is unfortunate, because cognitive values shape ML systems deeply. This process is replete with difficulties requiring extensive human deliberation, well beyond concerns about the opacity of internal representations. In particular, deliberation on cognitive values require data scientists going through two types of procedures, which I call value specification and value choice[9].

Let me start with *value specification*. The criteria we value in the different phases of training algorithms are usually vague and they can be specified in radically different ways. Take a straightforward valuable characteristic of data sets used to train algorithm, namely *representativeness*. This value can be conceived along different lines cutting social, ethnic, and technical factors, and depending on which factors you deem important, the data sets that are processed and used to train algorithms will look rather differently (Ratti and Graves 2022). Another example is related to how, for instance, we measure the performance of ML systems. Let's say we look at 'resources used' (e.g. computational), which is often mentioned as another important valuable thing that goes into evaluating performances. In other words, one can say that we value systems that use resources parsimoniously. But parsimony is difficult to specify. For instance, Lindauer and Hutter (2019) mentions that there are at least three different metrics that can be used to measure the 'resources used': there are wallclock-time, floating point operations, and epochs. If we opt for predictive power, then we have a similar problem. Consider k-nearest neighbors (kNN) algorithm and how it is evaluated by using Fbeta-measure. This metric is calculated by using precision (i.e. the number of correct positive predictions made out of all positive predictions) and recall (i.e. the number of correct positive predictions made out of all positive predictions that should have been made). However, the measure is obtained by deciding how to treat precision and recall, which is not obvious. Intuitively, maximizing precision will minimize false positives, while maximizing recall will minimize false negatives. In the tradition of inductive risk (Douglas 2009, Johnson 2021), how serious are the consequences of making a mistake in classifying a data point as *x* (e.g. *'tumor'*) rather than *y* (e.g. *'non-tumor'*) can determine our treatment of precision and recall, and hence of Fbeta-measure. This is yet another example of a point that should be well-taken: there is a significant amount of wiggle-room for specifying the desirable characteristics of ML systems (i.e., their 'values').

---
[8] (Hancox-Li and Kumar, 2021) and (Birhane et al, 2020) are two exceptions
[9] Value specification and value choice reflects the problems of theory choice that Kuhn discusses (1977), and that makes him tend towards the idea that theory choice is a value-laden process.

The second type of deliberation happens at the level of *value choice*: criteria like values are usually in tension or in non-specified relations one with the other, and choosing which value to promote is unclear. We can value *data quality*, but sometimes this value may stay in a tradeoff relation with, say, *representativeness*. For instance, in preparing representative data sets in a medical context, these may come from an underserved area, especially if we interpret representativeness as cutting across specific social factors, with the result of processing very poor data with lots of gaps. In this case, sticking with the value of representativeness comes at the expense of data quality. However, things may go also the other way around, in the sense that we may deem *data quality* as too important to be sacrificed for representativeness.

To point to the bigger picture, value-laden choices are an important component of the 'epistemic perspective' of ML systems. Through these choices specific interests, methods, presuppositions, problems and questions, hypotheses, theories, etc are integrated and stabilized in a single epistemic perspective characterizing a given ML system. This shows that ML systems are far from being non-human: they reflect the values of the human beings who design them. Specifying and choosing the 'virtues' of ML systems – no matter how opaque their internal representations are - requires plenty of value-laden judgements that bring back human agency to the epistemological center of ML-based science, and put scientists in a position of at least partial epistemic control. The way these choices are exercised are relevant for tracking conditions: a ML system is not just its internal representations, but also other components that are evaluated on the basis of clear criteria that are responsive to relevant epistemic and methodological standards, and that are established by means of value specification and value choice. Tracing conditions also apply: ML practitioners, in collaboration with other practitioners, continuously monitor the behavior of components of ML systems by appealing to specific methodological and epistemic criteria, which are the result of value specification and value choice dynamics.

## 3.3 The Values of Human Scientists and the Values of Machine Learning

Despite this promising strategy, it is important to point out epistemic control is not a one-way dynamic, flowing from human scientists to ML systems. A comprehensive account of ML-based science, while including human values as fundamental ingredients in shaping ML systems, cannot ignore that these systems – taken in isolation - bring their own perspectivity too. Humphreys has conceptualized this perspectivity in terms of representational style, and explained why this impairs epistemic control. However, even taken in isolation and separated from human beings, there is much more to the epistemic perspective of ML than just the 'representational style' of its models. What I claim here is that there is indeed something in ML-based science that shapes human epistemic control, but it is not (only) its style of representation. Here I claim that ML systems have values too, and this can constrain or be in tension with human values. However, it is notoriously difficult to argue that a technical artifact can embed or contain values (Pitt 2014), which is something more controversial than merely saying that they are value-laden. What I say instead is that a ML system brings its own normativity, and this constrains epistemic control because the 'standards' or 'criteria' necessary for tracking and tracing conditions are in part imposed by ML systems themselves. The notion of 'normativity' that I employ comes from Radder (2019), and I adapt it to fit scientific practice rather than just technology (as he does).

According to Radder (2019), a norm is an embedded directive "concerning what people should (and should not) say or do" (pp 45-46). The nature of norms presupposes value judgements on the desirability of certain actions, and they are, to a certain extent, prescriptive. For this reason, norms imply values. Norms vary in their nature, from epistemic to moral. The focus of Radder's work is on how technologies can be normative. For the sake of this article, the normativity of a given technological artefact can be seen as the fact that "its realization implies one or more norms or normative claims about what to say or do" (Radder 2019, pp 57-58). Technologies can be normative because their stable functioning in a region of space and time "requires that the people in that region should behave in such a way as to enable, and not disturb, the intended functioning of the technology" (p 58).

This general sense of normativity applies also to ML systems[10]. To paraphrase Radder, if we want to fully practice ML-based science, then conditions for its successful realization should be satisfied. But in satisfying these conditions, human beings necessarily have to assume certain characteristics of (good) science as given. Accepting these characteristics is, in a sense, inevitable in case we want ML-based science to function properly. This will push human scientists to accept certain implicit norms about scientific practice that can be imposed by the fact that ML systems have certain characteristics and not others. These norms lie at several levels, and most important at the level of cognitive values informing how science ought to be practiced. Because my space is limited, I will just show how this dynamics unfolds with respect to *value* choice.

---

[10] Interestingly, it applies to any scientific methodology

      Consider, again, the case of cancer genomics. As I explained above, ML models in this context are so complex, that no one can build a how-actually mechanistic model out of them. Most notably, what users can do with these ML models is disease stratification and classification at the molecular level. In this case, a characteristic of ML systems (opacity in the sense specified above) constrains the aims of molecular biologists. While molecular biologists see mechanistic explanations as the goal of their discipline (and, with that, the epistemic values of good mechanistic explanations), the use of ML tools guides the practice of science towards other goals (e.g. prediction and classification), thereby 'constraining' which epistemic values biologists will have to prioritize. In this case, the values of predictive models ought to be prioritized at the expense of the virtues of mechanistic explanations (Craver and Darden 2013). In other words, while a community of scientists (e.g. molecular biologists) may promote certain values and aims of science, ML-based science may dictate its own values because of the way ML tools work. Think about this, again, in terms of tracking: the system is responsive to methodological and epistemic criteria, but it is the system itself that tells us which criteria, among the many we have available, it should be responsive to.

      The normative impact of ML-based automated science on value specification and value choice implies that human epistemic control can be in principle only partial. Via value specification and value choice, scientists can standardize how ML systems are built, and have them to be responsive to discipline-specific conditions negotiated in scientific communities. At the same time, the way ML tools work poses serious constraints on what can be standardized, controlled, and to what aims. The big picture of epistemic control in ML-based is therefore one in which scientists have to come to terms with the characteristics of ML. Scientific norms are shaped and evolve as a result.

## 4. CONCLUSION

In this article, I have investigated the relation between ML-based science and the idea of meaningful human control. First, I have defined what is meaningful human control in the scientific context. I called this 'epistemic control', and I have defined it in terms of tracking and tracing conditions. Next, I have analyzed Humphreys' view about ML-based science, and I have translated his concerns in the language of epistemic control and tracking/tracing conditions. But an analysis of the practice of ML-based science reveals that his pessimism about epistemic control is unwarranted. Humans' perspective is still central because of the way cognitive values shape algorithmic systems. At the same time, characteristics of ML-based science have a normative impact for value specification and value choice, and such an impact shapes epistemic control in important ways. This means that ML tools should be used with cautious in science. The hype and the excitement surrounding ML methodologies can easily obfuscate the fact that these tools are likely to invisibly push scientific communities towards certain scientific aims and neglect others. This is already well underway in biology (Lopez-Rubio and Ratti 2021; Lemberger 2024), and it may as well be in other natural sciences.

As a closing note, it is also interesting to reflect on two things. First, the normative impact of ML-based science is not an outlier – all scientific methodologies likely have this impact. Second, to my knowledge this normative impact has not received much attention in the philosophy of science, with few exceptions. Literature on inductive and epistemic risks (Douglas 2009; Biddle and Kukla 2017) tends to represent idealized value deliberations, where scientists can in principle shape scientific methodologies to realize whatever value they have in mind. However, how scientific methodologies (metaphorically) respond to human values has not received much attention, and neglecting this aspect may have serious repercussions in terms of epistemic misalignments (Suarez and Boem 2022; Ratti and Russo 2024).